# Hybrid access control with modified SINR association for future heterogeneous networks

Aleksandar Ichkov, Vladimir Atanasovski and Liljana Gavrilovska
Ss. Cyril and Methodius University in Skopje, Macedonia
Faculty of Electrical Engineering and Information Technologies
{ichkov, vladimir, liljana}@feit.ukim.edu.mk

*Abstract*— The heterogeneity in cellular networks that comprise multiple base stations' types imposes new challenges in network planning and deployment of future generation of cellular networks. The Radio Resource Management (RRM) techniques, such as dynamic sharing of the available resources and advanced user association strategies, determine the overall network capacity and efficiency. This paper evaluates the downlink performance of a two-tier heterogeneous network (consisting of macro and femto tiers) in terms of rate distribution, i.e. the percentage of users that achieve certain rate in the system. The paper specifically addresses the femto tier RRM by randomization of the allocated resources and the user association process by introducing a modified SINR association strategy with bias factor for load balancing. Also, the paper introduces hybrid access control mechanism at the femto tier that allows the authorized users of the femtocell, which are part of the Closed Subscriber Group (CSG) on the femtocell, to achieve higher data rates up to 10 times compared to the other regular users associated in the access.

I. INTRODUCTION

The continuing growth in users' demands for better mobile broadband leads the development towards 5G [1-7] fostering new technologies and concepts [8]. One of the key novel 5G paradigms is the *network densification* [8] aimed at addressing the increased need for capacity. The densification can be achieved either in space, by increasing the number of network nodes in the system, or in frequency, by utilizing different portions of spectrum in different bands.

The space densification is a prerequisite for reducing the load factor at different network nodes and enhancing the overall network performances through the received signal power per user. It can be achieved by adding new Macro Base Stations (MBSs), which is an expensive solution not able to solve the problem of indoor coverage, or by adding additional pico/femto tier of base stations that differ from the macro tier in terms of transmission power, capacity and base station spatial density, resulting in network heterogeneity. The deployment of these small base stations is inexpensive and uncoordinated, which makes them preferable for coverage both indoor and outdoor, and data rate improvement.

The introduced network heterogeneity with the second tier of small base stations (macro-femto) requires advanced Radio Resource Management (RRM) techniques [9]. The femto tier is usually deployed sporadically and randomly without any spatial or frequency planning [10]. Additionally, the deployment of the Femto Base Stations (FBSs) is usually uncoordinated with the macro tier, which further complicates the design of intelligent strategies for resource allocation and sharing. For instance, the interference in the heterogeneous network can be classified as either *cross-tier* interference, caused by elements from different tiers (e.g. from a femtocell to a macrocell) or *co-tier* interference, between elements of the same tier (e.g. between neighboring femtocells). Cross-tier interference is easily alleviated in an orthogonal deployment of macrocells and femtocells at the expense of decreasing the spectral efficiency of the network. However, co-channel deployments, where carriers are shared between both tiers, can result in higher spectral efficiency if appropriate intereference cancelation is applied [11].

Another important aspect for the overall network performance and resource utilization is the user-to-BS association strategy. Traditional user-to-BS association strategies, such as Nearest BS association and Maximum Received Power association, are mostly BS coverage based and favor the MBSs for user association. This leads to severe increase of the interference at the femtocells, as the high power transmission from the macrocells greatly shrinks the femtocell's coverage areas, leading to a high underutilization of femtocell's physical resources.

This paper evaluates the performances of a two-tier LTE network with *randomization* of the allocated resources at the femto tier. We introduce a *modified SINR association strategy with biasing handoff factor*, which is used proactively in the user association phase to improve the rate distribution and the load factors on both tiers. Furthermore, the paper introduces a *hybrid access control mechanism* that allows higher data rates to be guaranteed for the subscribers of the femtocells compared to other associated users in the network.

The paper is organized as follows. Section II elaborates the possible access control mechanisms for heterogeneous LTE networks. Section III introduces a novel system model for performance analysis, whereas section IV introduces novel association strategy. Section V provides performance results for the introduced system. Finally, section VI concludes the paper.

II. ACCESS CONTROL MECHANISMS

The access control to the femto tier is a necessity in heterogeneous LTE networks because of potential signaling overhead burden. If all network users are allowed to connect to all FBSs, then the number of handovers will dramatically increase and the interference mitigation would be very challenging. Therefore, access control mechanisms play an important role and have to be carefully chosen depending on the network scenario and the user profile.

Different approaches have been proposed for the access control mechanisms [12]:

- *Open access*: all customers of the operator's network have the right to make use of any of the available femtocells.
- *Closed access*: only a subset of the users (defined by the femtocell owner) can connect to the femtocell. This model is referred to as Closed Subscriber Group (CSG).
- *Hybrid access*: can be used to compromise between authorized and non-authorized users as a limited amount of the femtocell resources are available to all users, while the rest are operated in a CSG manner. This paper proposes a specific hybrid access control mechanism detailed in section IV.

To successfully implement the access control mechanism in a network, the users are divided into two categories so that the femtocell is able to distinguish the authorized users that have privileges over its resources from the others:

- *Subscriber*, referring to a user that is registered in the femtocell's CSG, also known as authorized user.
- *Non-subscriber*, referring to a user that is not registered in the femtocell's CSG.

Different access control mechanisms have different performances depending on the underlying network and traffic model. The following section introduces a system model that will be later used for simulation analysis of a hybrid access control with novel association metric for the users.

### III. TWO TIER NETWORK MODEL

Our network represents a heterogeneous network model comprising of two types of network nodes, macro and femto base stations, deployed over an area $A$, as shown at Fig. 1.

The macro tier BSs, MBSs, are distributed using regular MBSs distribution, in a grid, so that the targeted area is divided into hexagonal cells with each of the MBSs located at the center of a hexagonal cell. The actual number of deployed MBSs needed to cover the targeted area is denoted with $N_m$, with distance $d$ between two adjacent MBSs. The downlink transmit power of each MBS is denoted as $P_m$.

The femto tier BSs, FBSs, are randomly distributed over the same area, uncoordinated with the macro tier, modeled by Poisson Point Process (PPP), as one of the most frequently used statistical distribution for modeling stochastic, two-dimensional point processes [9]. The intensity of the femto tier PPP is denoted as $\lambda_f$, with an average number of FBSs, $N_f = \lambda_f A$. The downlink transmit power of the FBSs is much lower than the one of the MBSs, $P_f << P_m$.

The users are also distributed according to PPP with intensity $\lambda_u$ and an average number $N_u = \lambda_u A$.

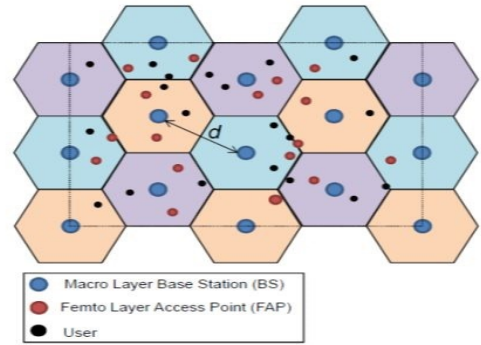

Figure 1. Network deployment over the observed area $A$

#### A. Radio Resource Management and user resource allocation

The radio access technology used is OFDMA, where the smallest resource unit that can be allocated to a single user is referred to as a Physical Resource Block (PRB) and denoted as $W_{PRB}$ (it occupies a certain frequency bandwidth and several time slots). The macro tier uses Hard Frequency Reuse (HFR) and each MBS gets the whole available bandwidth at its disposal with $N_{PRB,m} = N_{PRB}$ (the frequency reuse factor is assumed to be 1). The femto tier employs uncoordinated approach for resource allocation in order to mitigate co-tier interference and alleviate cross-tier interference, as it requires minor coordination with the macro tier to determine the overall network traffic state. The available system bandwidth is divided into $n_f$ spectrum fragments, each consisting of $N_{PRB,f} = N_{PRB}/n_f$ PRBs, assigned to every FBS randomly.

User resource allocation strategy requires each BS to distribute the available physical resources *fairly*, using equal power allocation to each PRB, so that each user gets the maximal possible data rate with respect to the network settings and traffic load.

### IV. USER-TO-BS ASSOCIATION FOR DOWNLINK TRANSMISSION

The user-to-BS association strategy determines which BS the user associates to for downlink transmission. The common approach is that each user associates to the BS, macro or femto, that maximizes a predefined association rule.

Let $M$ denote the set of all MBSs, $F$ the set of all FBSs and $V = M \cup F$ denote the set of all BSs, both macro and femto, in the network. Each user associates with the BS $k$, in accordance with the following general association rule:

$$k = \underset{i \in V}{argmax} \left\{ T_i Z_i^{-\gamma} \right\} \quad (1)$$

where $Z_i$ is the distance between the $i^{th}$ BS and the user, $\gamma$ is the path loss exponent and $T_i$ is referred as association weight. For example, if $T_i >> T_j$, $i \in F$, $j \in M$, then more traffic is routed on the femto tier. By adjusting $T_i$, the system can control the distribution of traffic on both tiers.

Existing user association strategies used in heterogeneous networks deployments depend on the value of the association weight $T_i$:

- *Nearest BS association* (MBS or FBS): $T_i = 1, \forall i \in V$
- *Cell range modification*: $T_i = P_i B_i$ and the range is extended if the bias factor $B_i > 1$ or reduced if $B_i < 1$.
- *Femtocell range extension*: if the bias factor $B_i > 1, \forall i \in F$ and $B_i = 1, \forall i \in M$
- *Maximum received power association*: if the bias factor $B_i = 1, \forall i \in V$ then $T_i = P_i, \forall i \in V$

In most cases, the general user association rule (1) results in high number of users being associated to the MBSs. This is critical in congested networks, where there are high number of denied users by the macro tier and high portions of the physical resources of the femto tier unused. Forcing more traffic routing to the femto tier might also result in congested FBSs, depending on the number of available physical resources on the femto tier. Therefore, there is a need for implementation of a load balancing between the different tiers of the networks to minimize the number of rejected users.

*A. Modified SINR User-to-BS association*

This paper introduces a *modified SINR user-to-BS association strategy* that adds a *load-balancing factor* in a maximum SINR association strategy to balance the amount of traffic routed at both tiers. It is enabled by the use of a *bias factor b* to modify the transmit power factor of the FBSs in the association phase, expanding the femtocell coverage area to force more users to associate with the FBSs.

The modified SINR association strategy is based on the association rule (1) with association weight $T_i$ defined as:

$$T_i = \frac{H_i(P_{i,prb} + b)}{\sum_{\substack{j \neq i \\ j=1 \\ j \in M}}^{N_m} H_j P_{j,prb} + \sum_{\substack{j \neq i \\ j=1 \\ j \in F}}^{N_f} H_j(P_{j,prb} + b) + N_0} \quad (2)$$

where $H_i$ defines the Rayleigh channel coefficient for the $i^{th}$ user, $P_{i,prb}$ defined the downlink transmit power on a single PRB of the targeted BS defined as $P_m / N_{PRB,m}$ for the MBSs and $P_f / N_{PRB,f}$ for the FBS and b denotes the bias factor. The first sum in the denominator denotes the interference from the MBSs in the system, while the second sum denotes the interference from the FBSs in the system. The last element in the denominator $N_0$ denotes the noise power.

Note that, if the bias factor *b* is equal to zero for all the MBSs and FBSs, then the association strategy corresponds to the Maximum SINR association, with a normalized transmit power factor $P_{i,prb}$. The Maximum SINR association strategy associate the users to the BS that maximizes the SINR value, which favors the MBSs because of the much higher transmit power compared to the FBSs. It maximizes the average data rate per user, but leads to higher number of rejected users on the macro tier and underutilization of the femto tier resources.

Through the usage of different values for the bias factor b, we can balance the load factor on the femto tier and avoid underutilization of the femto tier resources. The power factor $P_{i,prb}$ used in (2) depends on the number of PRBs available to the FBS. If the number of PRBs per FBS is the same as the number of PRBs per MBS, we need a larger bias factor to balance for the higher values for $P_{i,prb}$ on the MBSs due to the higher transmit power. That way, we are able to precisely control the amount of traffic routed to the femto tier to optimize the load factors on both tiers. This improves the femto resource utilization and decreases the number of rejected users in the network, compared to the maximum SINR association. As a trade-off, the average data rate per user decreases, because some users may not associated with the corresponding BS that maximizes their SINR value.

*B. Hybrid access control on femto tier*

We also introduce a modified hybrid access control that combines the advantages of both open and closed access control, where all femtocell resources are available to all users, with a preferential access for the subscribers using higher priority compared to non-subscribers. In our scenario, the list of subscribers, CSG, is comprised of 5 users randomly chosen from the femtocell's coverage area. In case there is at least one subscriber's association request at the femtocell, than all available resources are fairly allocated among the subscribers, while the other non-subscriber's requests are rejected. If there are no subscribers' requests at the femtocell, the femtocell allows association with the non-subscribers, fairly allocating all resources among them.

In this sense, the hybrid access control enhances the resource utilization in comparison to closed access, keeping the preferential access of the femtocell's resources for the subscribers. Another advantage is that all femtocell's resources are utilized, compared to the previously detailed hybrid access control mechanism that allows only a limited amount of the femtocell resources to be used by the non-subscribers, while the rest are available only to the subscribers.

V. PERFORMANCE EVALUATION

In order to evaluate the performance of our two-tier cellular network, the proposed network model is simulated using MATLAB. The simulation parameters used are given in Table I.

| Table I. Simulation parameters | |
|---|---|
| Observed area ($A$) | 25 km² (5x5) |
| Total system bandwidth | 20 MHz (100 PRBs) $W_{PRB}$ = 180 kHz |
| Macro BS transmit power | 43 dBm |
| Femto BS transmit power | 20 dBm |
| Distance between two macro BSs | 1 km |
| Number of fragments for femto tier bandwidth randomization | 1, 2, 4 and 10 with 100, 50, 25 and 10 PRBs respectively |
| Wireless channel gain distribution | Rayleigh (unit var.) |
| Path loss exponent | 2,3 |
| Noise power | $10^{-12}$ |
| Number of Macro BSs | 33 |
| Average number of FBSs | 200 |
| Average number of users | 5000 |
| Radius of femtocell coverage | 200 m [13] |
| Number of subscribers per FAP | 5 |

The SINR calculation in OFDMA based networks requires incorporation of the number of scheduled PRBs for a particular user and the number of overlapping PRBs with other MBSs or FBSs in the area. Therefore, we introduce SINR calculation formula that uses novel approach to calculate the interference from surrounding BSs with overlapping PRBs:

$$SINR_{ij} = \frac{\frac{\alpha_{ij}P_j}{N_{PRB,j}}\|h_{ij}\|^2 \|x_{ij}\|^{-\alpha}}{\sum_{\substack{m=1 \\ m \neq j \\ j \in M}}^{N_m} \frac{\beta_{im}P_m}{N_{PRB,m}}\|h_{im}\|^2 \|x_{im}\|^{-\alpha} + \sum_{\substack{f=1 \\ f \neq j \\ j \in M}}^{N_f} \frac{\beta_{if}P_f}{N_{PRB,f}}\|h_{if}\|^2 \|x_{if}\|^{-\alpha} + N_0} \quad (3)$$

Equation (3) represents the received SINR at the $i^{th}$ user that is associated to the $j^{th}$ BS ($j$ is either in $M$ or $F$). $\alpha_{ij}$ is random variable that represents the number of PRBs allocated to the user. The total transmit power from the $j^{th}$ base station to the $i^{th}$ user is $\alpha_{ij}P_j/N_{PRB,j}$, where $P_j/N_{PRB,j}$ is the transmit power on one RB from the $j^{th}$ base station. $h_{ij}$ and $x_{ij}$ are the channel fading and the distance between the $i^{th}$ user and the $j^{th}$ BS, respectively. The first sum in the denominator denotes the interference from MBSs in the system, while the second sum denotes the interference from FBSs in the system. The random variables $\beta_{im}$ and $\beta_{if}$ represent the number of overlapping PRBs between the $i^{th}$ user and the interfering BSs, both macro and femto, respectively. The parameter $N_0$ is the noise power.

The rate $R$, for the $i^{th}$ user in the system, with respect to the received SINR, is calculated as:

$$R_{ij} = \alpha_{ij}W_{PRB}\log_2\left(1+SINR_{ij}\right) \quad (3)$$

The rate distribution $\Psi$ for the association users, or equivalently the probability that certain percentage of users achieve rate higher than a predefined threshold, is defined as:

$$\Psi = \Pr[R > \delta \mid R > 0] \quad (4)$$

The goal is to maximize the average rate in the system that can be guaranteed to any associated user.

*A. Simulation results*

The simulation scenarios analyze the rate distribution for different number of femto tier spectrum fragments. We use the modified SINR association strategy in the user association phase, with different values for the bias factor according to the sizes of femto tier spectrum fragments.

Figure 2 shows the rate distribution for the two-tier network, in comparison to the number of femto tier spectrum fragments and the corresponding bias factors used in the modified SINR association. As we can see, there is a trade-off between attaining high data rates per user and the percentage of users guaranteed to achieve those rates. If the operator targets high data rate for small percentage of users, then the femto tier should use larger spectrum fragments. However, if the operator wants to guarantee a predefined, lower data rate to higher number of users, the femto tier should use smaller spectrum fragments, which efficiently mitigates the inter/intra tier interference in congested network, but results in more rejected users. With the implementation of load balancing, more traffic is routed on the femto tier which increases the number of associated users on the femtocells, as shown at Fig. 2b. That improves the utilization of femto tier resources, compare to the case when load balancing is not implemented, where a large portion of the femto resources remain underutilized due to the users' tendency to associate with high power MBSs, resulting in high number of rejected users.

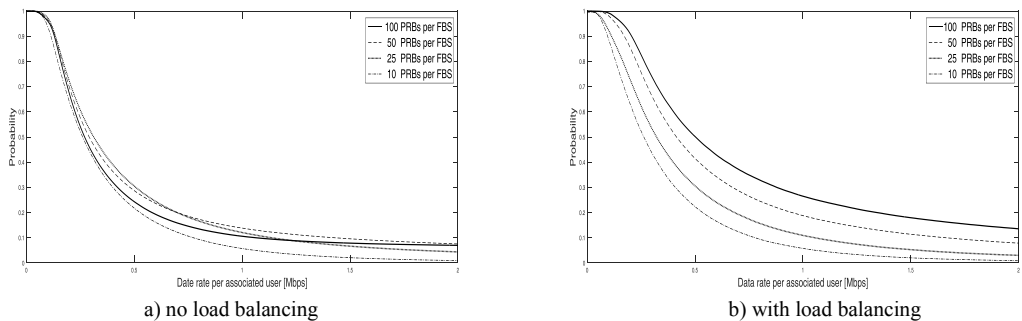

a) no load balancing      b) with load balancing

Figure 2. Rate distribution $\Psi$ using modified SINR association strategy

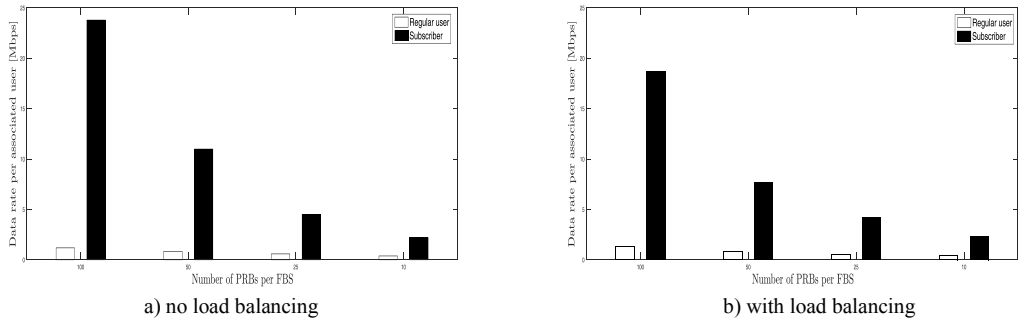

a) no load balancing  b) with load balancing

Figure 3. Average date rate per associated user vs. subscriber

Therefore, with the regulation on the bias factor values in accordance with the available femto tier resources, we can guarantee a certain data rate for higher number of users and decrease the number of rejected users in the network.

Next we compare the average data rates achieved for the associated users in the network using the same network settings. Fig. 3a shows the average rates per associated user and subscribers in the network, using the modified SINR association strategy with no load balancing implanted while Fig. 3b shows the average rates per associated user and subscribers in the network with implemented load balancing. As the results shows, the embedded hybrid access control at the femto tier enables the subscribers to achieve data rates up to 10 times higher than the average data rate for all associated users in the networks, due to the subscribers' authorized access on the femtocell's resources. The average rates per associated users are lower when there is load balancing implemented in the network due to the higher number of associated users and the limited physical resources available in the network.

Co-channel scenarios, without any explicit interference management, typically use small values of the bias factor (e.g. few dBs) as femtocell users will otherwise experience too high interference from the macro layer [14]. An autonomous FBS power setting schemes is introduced in 3GPP Rel. 10, where the FBS transmit power is adjusted to minimize the interference generated for nearby macro users. Also a new enhanced ICIC scheme for heterogeneous OFDMA networks has been developed which offers time-domain resource partitioning between network layers for better performance and will allow using higher bias factor values.

## VI. CONCLUSIONS

This paper analyzes the spectrum resource allocation, sharing and utilization efficiency in a two-tier OFDMA heterogeneous network with macro and an additional uncoordinated hybrid controlled femto tier. The femto tier uses simple spectrum fragmentation and random fragment allocation to determine the operating resources. The results show that the rate distribution of the system depends on the fragment size for varying network conditions and suggests that the femto tier can dynamically adjust it, according to the network congestion state. A modified SINR association strategy is used to enable the network to balance the traffic routed on both tiers and decrease the number of service denied users in the network.

From operators' perspective, the results can also be used to implement an intelligent RRM, where the fragment size at the femto tier can be dynamically adjusted according to the congestion in the network, by providing loose coordination with the macro tier, with optimized load balancing factor depending on the appropriate fragment size to fully utilize all available femto tier resources. The embedded hybrid access control mechanism at the femtocells allows higher data rates to be guaranteed for the authorized users of the femtocell, up to 10 times compared to the other users in the networks. At the same time, the operator can utilize the hybrid femtocell's resources when there are no authorized users' requests. This can be used to motivate the usage of femtocells, especially in indoor environments such as homes, offices etc.

## ACKNOWLEDGMENT

This work was supported by the Public Diplomacy Division of NATO in the framework of Science for Peace through the SfP-984409 ORCA project and by the EC FP7 eWall project (http://ewallproject.eu/) under grant agreement No. 610658.